\documentstyle[psfig]{mn}
\def\etal{{\rm et al.}}

\def\P3M{P$^3$M}
\def\refindent{\par \noindent \hang}
\def\paper#1#2#3#4#5{\refindent #1, #2, #3, #4, #5}

\def\and{, }
\begin{document}

\title[Gravitational lensing in sCDM] {Weak gravitational lensing in
the standard Cold Dark Matter model, using an algorithm for
three-dimensional shear}

\author[Andrew J. Barber \etal] {Andrew J. Barber$^1$\thanks{Email:
abarber@star.cpes.susx.ac.uk}, Peter A. Thomas$^1$ and
H. M. P. Couchman$^2$ \\
{}$^1$Astronomy Centre, University of Sussex, Falmer, Brighton, BN1
9QJ \\
{}$^2$Dept.~of Physics \& Astronomy, Univ.~of Western Ontario,
London, Ontario, N6A 3K7, Canada 
}

\date{Accepted 1998 ---. Received 1998 ---; in original form 1998 ---}

\maketitle

\begin{abstract}

We investigate the effects of weak gravitational lensing in the
standard Cold Dark Matter cosmology, using an algorithm which
evaluates the shear in three dimensions. The algorithm has the
advantage of variable softening for the particles, and our method
allows the appropriate angular diameter distances to be applied to
every evaluation location within each three-dimensional simulation
box. We investigate the importance of shear in the distance-redshift
relation, and find it to be very small. We also establish clearly
defined values for the smoothness parameter in the relation, finding
its value to be at least 0.88 at all redshifts in our
simulations. From our results, obtained by linking the simulation
boxes back to source redshifts of 4, we are able to observe the
formation of structure in terms of the computed shear, and also note
that the major contributions to the shear come from a very broad range
of redshifts. We show the probability distributions for the
magnification, source ellipticity and convergence, and also describe
the relationships amongst these quantities for a range of source
redshifts. We find a broad range of magnifications and ellipticities;
for sources at a redshift of 4, 97\mbox{$\frac{1}{2}$}\% of all lines
of sight show magnifications up to 1.3 and ellipticities up to
0.195. There is clear evidence that the magnification is not linear in
the convergence, as might be expected for weak lensing, but contains 
contributions from higher order terms in both the convergence and the 
shear.

\end{abstract}


\section{INTRODUCTION}

The gravitational lensing of light by the general form of the
large-scale structure in the universe is of considerable importance in
cosmology. This `weak lensing' may result in magnification of a
distant source from Ricci focusing due to matter in the beam, and
shear leading to distortion of the image cross-section. The strength
of these effects depends on the lens and source angular diameter
distances and the specific distribution of matter between the observer
and source. Consequently the effects are likely to be sensitive to the
particular cosmological model. In extreme cases, a source may be
strongly lensed if the light passes close to a massive structure such
as a galaxy, and this occasionally results in the appearance of
multiple images of the source. One of the most important applications
of such `strong lensing' studies has been the reconstruction of mass
profiles for lensing galaxies and estimations of the Hubble parameter,
$H_0$, from measurements of the time-delay between fluctuations in the
multiple images of a background quasar; see, e.g., Falco, Govenstein
and Shapiro (1991), Grogan and Narayan (1996), and Keeton and Kochanek
(1997). These studies have frequently made use of the `thin-screen
approximation' in which the depth of the lens is considered to be
small compared with the distances between the observer and the lens
and the lens and the source. In the thin-screen approximation the mass
distribution of the lens is projected along the line of sight and
replaced by a mass sheet with the appropriate surface density
profile. Deflections of the light from the source are then considered
to take place only within the plane of the mass sheet, making
computations for the light deflections much simpler.

The simplicity of the thin-screen approximation has also lead to its
use in weak gravitational lensing studies, where the output volumes
from cosmological $N$-body simulations are treated as planar
projections of the particle distributions within them. However, the
procedures have to be extended when dealing with the propagation of
light from very distant sources, where a number of simulation time
outputs are necessary to cover the observer-source distance. In these
cases each simulation volume is replaced by a planar projection of the
particle distribution, and to compute the distributions in
magnification and shear for a large number of rays passing through the
system of screens, use is made of the multiple lens-plane theory which
has been variously described by Blandford and Narayan (1986),
Blandford and Kochanek (1987), Kovner (1987), Schneider and Weiss
(1988a, b), and summarised by Schneider, Ehlers and Falco (1992). We describe
some of these two-dimensional weak lensing methods in Section 1.1.

Couchman, Barber and Thomas (1998) considered some of the shortcomings
of these two-dimensional lens-plane methods, and also rigorously
investigated the conditions under which two-dimensional methods would
give equivalent results to integrating the shear
components\footnote{Note that, throughout this paper we refer to the
elements of the matrix of second derivatives of the gravitational
potential as the `shear' components, although, strictly, the term
`shear' refers to combinations of these elements which give rise to
anisotropy.} through the depth of a simulation volume. They showed
that, in general, it is necessary to include the effects of matter
stretching well beyond a single period in extent, orthogonal to the
line of sight, but depending on the particular distribution of
matter. It is also necessary to project the matter contained within a
full period onto the plane, assuming the distribution of matter in the
universe to be periodic with periodicity equal to the simulation
volume side dimension. They also showed that errors can occur in
two-dimensional approaches because of the single angular diameter
distance to each plane, rather than specific angular diameter
distances to every location in the simulation volume.

These considerations motivated Couchman et al. (1998) to develop an
algorithm to evaluate the shear components at a large number of
locations within the volume of cubic particle simulation
time-slices. The algorithm they developed is based on the standard
P$^3$M method (as described in Hockney and Eastwood, 1988), and uses a
Fast Fourier Transform (FFT) method for speed. It is designed to
compute the six independent three-dimensional shear components, and
therefore represents a significant improvement over two-dimensional
methods. We describe the three-dimensional shear algorithm in outline
in Section 2.1.

In this paper we have applied the algorithm to the standard Cold Dark
Matter (sCDM) cosmological $N$-body simulations available from the
Hydra consortium,\footnote{ http://coho.astro.uwo.ca/pub/data.html}
which we describe in Section 2.2. By combining the outputs from the
algorithm from sets of linked time-slices going back to a redshift of
4, we are able to evaluate the overall shear, convergence,
magnifications and source ellipticities (and distributions for these
quantities). We first describe other work which has generated results
from studies of weak lensing in the sCDM cosmology.

\subsection{Other work}

There are numerous methods for studying weak gravitational lensing. In
`ray-tracing,' (see, for example, Schneider and Weiss, 1988b,
Jaroszy\'{n}ski et al., 1990, Wambsganss, Cen and Ostriker, 1998, and
Marri and Ferrara, 1998, the paths of individual light rays are
traced backwards from the observer as they are deflected at each of
the projected time-slice planes. The mapping of these rays in the
source plane then immediately gives information about the individual
amplifications which apply. In the `ray-bundle' method, (see, for
example, Fluke, Webster and Mortlock, 1998a, b, and Premadi, Martel
and Matzner, 1998a, b, c), bundles of rays representing a circular image are
considered together, so that the area and shape of the bundle at the
source plane, (after deflections at the intermediate time-slice
planes), gives the required information on the ellipticity and
magnification. There are also many different procedures for computing
the deflections and shear, although most apply the multiple lens-plane
theory to obtain the overall magnifications and distributions. We
shall describe briefly four works which have produced weak lensing
results in the sCDM cosmology.

Jaroszy\'{n}ski et al. (1990) use the ray-tracing method with
two-dimensional planar projections of the time-slices, and by making
use of the assumed periodicity in the particle distribution, they
translate the planes for each ray, so that it becomes centralised in
the plane. This ensures that there is no bias acting on the ray when
the shear is computed. Each plane is divided into a regular array of
pixels, and the column density in each pixel is evaluated. Instead of
calculating the effect of every particle on the rays, the pixel column
densities in the single period plane are used. They calculate the two
two-dimensional components of the shear (see Section 5 for the
definition of shear) as ratios of the mean convergence of the beam,
which they obtain from the mean column density. However, they have not
employed the net zero mean density requirement in the planes,
(described in detail by Couchman et al., 1998), which ensures that
deflections and shear can only occur when there are departures from
homogeneity. Also, the matter in the pixel through which the ray is
located is excluded. Their probability distributions for the
convergence, due to sources at redshifts of 1, 3 and 5, are therefore
not centralised around zero, and exhibit only limited broadening for
sources at higher redshift. They also display the probability
distributions for the shear and the corresponding distributions for
source ellipticity. The procedures used by Jaroszy\'{n}ski (1991) and
Jaroszy\'{n}ski (1992) are improved by the introduction of softening
to each particle to represent galaxies of different masses and radii
with realistic correlations in position. In these papers Monte Carlo
methods are used to study the effects of weak lensing on the
propagation of light through inhomogeneous particle distributions.

Wambsganss et al. (1998) also use the ray-tracing method
with two-dimensional planar projections of the simulation boxes, which
have been randomly oriented. Rays are shot through the central region
of 8$h^{-1}$Mpc $\times 8h^{-1}$Mpc only, (where $h$ is the Hubble
parameter expressed in units of 100 km s$^{-1}$ Mpc$^{-1}$), and the
deflections are computed by including all the matter in each plane,
allocated to pixels 10$h^{-1}$kpc $\times 10h^{-1}$kpc, covering one
period in extent only. The planes have comoving dimensions of
80$h^{-1}$Mpc $\times 80h^{-1}$Mpc. The computations make use of a
hierarchical tree code to collect together those matter pixels far
away, whilst the nearby ones are treated individually, and the code
assumes that all the matter in a pixel is located at its centre of
mass. By using the multiple lens-plane theory, they show both the
differential magnification probability distribution, and the
integrated one for 100 different source positions at redshift
$z_s=3.0$. One advantage of this type of ray-tracing procedure is its
ability to indicate the possibility of multiple imaging, where
different rays in the image plane can be traced back to the same pixel
in the source plane.

Premadi, Martel and Matzner (1998a) have improved the resolution of
their $N$-body simulations by using a Monte Carlo method to locate
individual galaxies inside the computational volume, and ensuring that
they match the 2-point correlation function for galaxies. They also
assign morphological types to the galaxies according to the
individual environment, and apply a particular surface density profile
for each. To avoid large scale structure correlations between the
simulation boxes, five different sets of initial conditions are used
for the simulations, so that the individual plane projections can be
selected at random from any set. By solving the two-dimensional
Poisson equation on a grid, and inverting the equation using a FFT
method, they obtain the first and second derivatives of the
gravitational potential on each plane. They also correctly ensure that
the mean surface density in each lens-plane vanishes, so that a good
interpretation of the effects of the background matter is made. Their
method uses beams of light, each comprising 65 rays arranged in two
concentric rings of 32 rays each, plus a central ray. To obtain good
statistical data they have run their experiment for 500 beams. They
show the average shear for a source at $z_s=5$ contributed by each of
the lens-planes individually, and find that the largest contributions
come from those planes at intermediate redshift, of order
$z$=1~-~2. Similarly, they find that the lens-planes which contribute
most to the average magnifications are also located at intermediate
redshifts. The multiple lens-plane theory then enables the
distributions of cumulative magnifications to be obtained, which are
shown to be broad and similar in shape for the sCDM and cosmological
constant models, although the latter model shows a shift to larger
magnification values.

Marri and Ferrara (1998) use a total of 50 lens-planes evenly spaced
in redshift up to $z=10$. Their mass distributions have been
determined by the Press-Schechter formalism (which they outline),
which is a complementary approach to $N$-body numerical simulations.
From this method they derive the normalised fraction of collapsed
objects per unit mass for each redshift. They acknowledge that the
Press-Schechter formalism is unable to describe fully the complexity
of extended structures, the density profile of the collapsed objects
(the lenses), or their spatial distribution at each redshift. They
therefore make the assumption that the lenses are spatially
uncorrelated and randomly distributed on the planes, and furthermore
behave as point-like masses with no softening. The maximum number of
lenses in a single plane is approximately 600, each having the
appropriate computed mass value. In their ray-tracing approach they
follow $1.85 \times 10^7$ rays uniformly distributed within a solid
angle of $2.8 \times 10^{-6}$sr, corresponding to a $420\arcsec \times
420\arcsec$ field. The final impact parameters of the rays are
collected in an orthogonal grid of $300^2$ pixels in the source
plane. Because of the use of point masses, their method produces very
high magnification values, greater than 30 for the sCDM
cosmology. They have also chosen to use a smoothness parameter
$\bar{\alpha} =0$ in the redshift-angular diameter distance relation
(which we describe in Section 3) which depicts an entirely clumpy
universe.

\subsection{Outline of paper}

In Section 2.1 we summarise the main features of the algorithm for
shear in three dimensions, which is detailed in Couchman et
al. (1998). Because the code is applied to evaluation positions within
the volume of $N$-body simulation boxes, we are able to apply specific
appropriate angular diameter distances to each location, which is not
possible with two-dimensional planar projections of simulation
boxes. We also note that the algorithm automatically includes the
effects of the periodic images of the fundamental simulation volume,
so that the results for the shear are computed for matter effectively
stretching to infinity. Included also is the net zero mean density
requirement, which ensures that deflections and shear may only occur
as a result of departures from homogeneity. In Section 2.2 we describe
the sCDM $N$-body simulations and how we combine the different output
time-slices from the simulations to enable the integrated shear along
lines of sight to be evaluated. Section 2.3 describes the variable
softening facility employed in the code, and our choice of a minimum
softening value, which may be given a realistic physical
interpretation.

Because we evaluate the shear at locations throughout the volume of
the simulation boxes, and because of some sensitivity (see Couchman et
al., 1998) of the results to the smoothness, or clumpiness, of the
matter distribution in the universe, we consider, in Section 3, our
choice of the appropriate angular diameter distances. We consider the
effects of shear on the angular diameter distance, and the sensitivity
of our results to the smoothness parameter,
$\bar{\alpha}$. Measurements of the particle clustering within our
simulations, which determines the variable softening parameter for use
in the shear algorithm, also enable a good definition for the
smoothness parameter to be made, and this is discussed.

In Section 4, we describe the formation of structure within the
universe as it evolves, in terms of the magnitudes of the shear
components computed for each time-slice. We see how the rms values of
the components vary with redshift, and also how the set of highest
values behave. We also identify, in terms of the lens redshifts, where
the significant contributions arise. Our conclusions are compared with
the results of other authors.

In Section 5, we describe in outline the multiple lens-plane theory,
with particular reference to our application of it. In Section 6, we
discuss our results for the shear, convergence, magnifications, source
ellipticities, distributions of these values, and relationships
amongst them. Section 7 summarises our findings, compares our results
with those of other authors for the sCDM cosmology, and proposes
applications of our method and results.

\section{THE ALGORITHM FOR THREE-DIMENSIONAL SHEAR, AND THE
COSMOLOGICAL SIMULATIONS}

\subsection{Description of the three-dimensional algorithm}

The algorithm we are using to compute the elements of the matrix of
second derivatives of the gravitational potential has been described
fully in Couchman et al. (1998). The algorithm is based on the
standard P$^3$M method, and uses a FFT convolution method. It computes
all of the six independent shear component values at each of a large
number of selected evaluation positions within a three-dimensional
$N$-body particle simulation box. The P$^3$M algorithm has a
computational cost of order $N{\rm log}_2 N$, where $N$ is the number
of particles in the simulation volume, rather than $O(N^2)$ for
simplistic calculations based on the forces on $N$ particles from each
of their neighbours. For ensembles of particles, used in typical
$N$-body simulations, the rms errors in the computed shear component
values are typically $\sim 0.3\%.$

In addition to the speed and accuracy of the shear algorithm, it has
the following features.

First, the algorithm uses variable softening designed to distribute
the mass of each particle within a radial profile which depends on its
specific environment. In this way we are able to set individual mass
profiles for the particles which enables a physical depiction of the
large scale structure to be made. We describe our choice of the
appropriate variable softening in Section 2.3.

Second, the shear algorithm works within three-dimensional simulation
volumes, rather than on planar projections of the particle
distributions, so that angular diameter distances to every evaluation
position can be applied. It has been shown (Couchman et al., 1998)
that in specific circumstances, the results of two-dimensional planar
approaches are equivalent to three-dimensional values integrated
throughout the depth of a simulation box, provided the angular
diameter distance is assumed constant throughout the depth. However,
by ignoring the variation in the angular diameter distances throughout
the box, errors up to a maximum of 9\% can be reached at a redshift of
$z=0.5$ for sCDM simulation cubes of comoving side
100$h^{-1}$Mpc. (Errors can be larger than this at high and low
redshift, but the angular diameter distance multiplying factor for the
shear values is greatest here for sources we have chosen at a redshift
of 4.)

Third, the shear algorithm automatically includes the contributions
of the periodic images of the fundamental volume, essentially creating
a realisation extending to infinity. Couchman et al. (1998) showed
that it is necessary to include the effects of matter well beyond the
fundamental volume in general (but depending on the particular
particle distribution), to achieve accurate values for the
shear. Methods which make use of only the matter within the
fundamental volume may suffer from inadequate convergence to the
limiting values.

Fourth, the method uses the `peculiar' gravitational potential, $\phi$,
through the subtraction of a term depending upon the mean
density. Such an approach is equivalent to requiring that the net
total mass in the system be set to zero, and ensures that we deal only
with light ray deflections arising from departures from homogeneity;
in a pure Robertson-Walker metric we would want no deflections.

\subsection{The sCDM large scale structure simulations}

We have chosen, in this paper, to apply the shear algorithm to the
sCDM cosmological $N$-body simulations available from the Hydra
consortium, and produced using the `Hydra' $N$-body hydrodynamics
code, as described by Couchman, Thomas and Pearce (1995). Each
time-slice from this simulation contains $128^3$ dark matter
particles, each of 1.2 $\times 10^{11}h^{-1}$ solar masses, with a CDM
spectrum in an Einstein-de Sitter universe, and has comoving box sides
of $100h^{-1}$Mpc. The output times for each time-slice have been
chosen so that consecutive time-slices abut, enabling a continuous
representation of the evolution of large scale structure in the
universe. However, to avoid unrealistic correlations of the structure
through consecutive boxes, we arbitrarily rotate, reflect and
translate the particle coordinates in each before the boxes are linked
together. We have chosen to analyse all the simulation boxes back to a
redshift of 3.9, a distance which is covered by a continuous set of 33
boxes (assuming the source in this case at $z_s=3.9$ to be located at
the far face of the 33rd box, which has a nominal redshift of
3.6). The simulations used have a power spectrum shape parameter of
0.25 as determined experimentally on cluster scales, (see Peacock and
Dodds, 1994), and the normalisation, $\sigma _8$, has been taken as
0.64 to reproduce the number density of clusters, according to Vianna
and Liddle (1996).

We establish a regular array of $100 \times 100$ lines of sight
through each simulation box, and compute the six independent shear
components at 1000 evenly spaced evaluation positions along
each. Since we are dealing with weak lensing effects and are
interested only in the statistical distribution of values, these lines
of sight adequately represent the trajectories of light rays through
each simulation box. It is sufficient also to connect each `ray' with
the corresponding line of sight through subsequent boxes in order to
obtain the required statistics of weak lensing. This is justified
because of the random re-orientation of each box performed before the
shear algorithm is applied.

\subsection{Variable softening}

The variable softening facility in the code allows each particle to be
treated individually as an extended mass, and the softening parameter
applied to each is chosen to be proportional to the distance, $l$, to
the particle's 32nd nearest neighbour. In this way the softening is
representative of the density environment of each particle. The
appropriate value of the parameter is determined using a different
smoothed particle hydrodynamics (SPH) programme. The shear algorithm
then works with the ratio of the chosen softening for each particle to
the maximum value (equivalent to the mesh dimension, which is defined
by the regular grid laid down to decompose the short- and long-range
force calculations), so that the parameter has a maximum value of
unity in the code.

Isolated particles are therefore assigned large softening values, and
are then not able to cause anomalous strong deflections. In
addition, this helps to ensure that more rays pass through regions of
softened mass rather than voids of negative density. Particles in
denser regions are assigned correspondingly smaller softening scales,
and are therefore able to cause stronger deflections. In the regions
of highest density we choose to set a minimum value for the softening,
to avoid interpolation errors in the code for very small separations,
and to introduce a physically realistic scale size to such particles.

In the sCDM simulation we have used, the minimum values for $l$ are of
order $10^{-3}$ in box units, and for a large cluster of 1000
particles this is comparable to the maximum value of the Einstein
radius for lenses up to a redshift of 4. (For our maximum source
redshift of 3.9, and for a lens of 1000 particles in our simulation,
the Einstein radius reaches a maximum of 0.11$h^{-1}$ Mpc, or 0.0011
box units, at a redshift of 0.52.)  Hence, by choosing a minimum for
the variable softening of this order, we would rarely expect to see
strong lensing. At the same time, this scale is approximately of
galactic dimensions, thereby giving a realistic interpretation to the
choice.

We have therefore set the minimum level to 0.001 in box units, and
allowed it to remain at a fixed physical dimension throughout the
redshift range of the simulations. Thus, we have set the value to be
0.001 for the $z=0$ simulation box, rising to 0.0046 in the earliest
simulation box at $z=3.6$.

Couchman et al. (1998) describes also the sensitivity of the
magnification distributions to the choice of minimum softening arising
from a single, assumed isolated, simulation box, and shows that the
results are insensitive to minimum softenings of 0.001 and 0.002,
apart from the peak magnification values, which occur only in limited
numbers of lines of sight. This is very useful because we can assume
that our results are likely to be little different from those using
the same minimum softening throughout, whilst keeping the value fixed
in physical size gives a credible interpretation to the softening.

\section{ANGULAR DIAMETER DISTANCES}

One of the advantages of being able to evaluate the shear components
at a large number of locations within the volume of each time-slice is
that we are able to apply the appropriate angular diameter distance
factors to each as part of the procedure to determine the
magnifications and ellipticities. The elements of the Jacobian matrix,
at each evaluation position,
\begin{equation}
\cal A = \left( \begin{array}{cc}
	1-\psi_{11}  & -\psi_{12} \\
	-\psi_{21}   & 1-\psi_{22}
	\end{array}
	\right),
\end{equation}
(from which the magnification may be derived at any point), contains
the two-dimensional effective lensing potentials which are related to
the computed three-dimensional shear through
\begin{equation}
\psi_{ij} = \frac{D_d D_{ds}}{D_s}.\frac{2}{c^2} \int\frac{\partial^2
\phi(z)}{\partial x_i \partial x_j}dz,
\end{equation}
where $D_d$, $D_{ds}$, and $D_s$ are the angular diameter distances
from the observer to the lens, the lens to the source, and the
observer to the source, respectively, and $c$ is the velocity of
light. (The factor $D_dD_{ds}/D_s$ may be written equivalently as
$cR/H_0$, where $R$ is dimensionless.) The integration is
along the line of sight. The angular diameter distance of the source
is defined to be the distance inferred from its angular size, assuming
Euclidean geometry, and in an expanding universe this distance becomes
a function of the redshift of the source. The angular diameter
distance also depends very much on the distribution of matter; for
example, excess matter within the beam causes it to become more
focussed, making the source appear closer than it really is. It is
therefore necessary to have available appropriate values for the
angular diameter distances for the particular distribution of matter
in the simulation data-set being investigated.

Schneider et al. (1992) summarise clearly the work of Dyer
and Roeder (1972, 1973) who made assumptions about the type of matter
distribution to obtain a second order differential equation for the
angular diameter distance in terms of the density parameter, $\Omega$,
for the universe, and the redshift of the source:
\begin{eqnarray}
\lefteqn{\left(z+1 \right) \left(\Omega z+1\right)
\frac{d^2D}{dz^2}+\left(\frac{7}{2}\Omega z + \frac{\Omega}{2} + 3
\right)\frac{dD}{dz}} \nonumber \\
 & & \hskip 1.0 in + \left(\frac{3}{2}\bar{\alpha}\Omega + 
\frac{\mid \sigma \mid^2}{(1+z)^5} \right) D = 0.
\end{eqnarray}
$\bar{\alpha}$ is the smoothness parameter, which is taken to be the
fraction of mass in the universe which is smoothly distributed, so
that a fraction $(1-\bar{\alpha})$ is considered to be bound into
clumps. $\sigma$ is the optical scalar for the shear, introduced by
matter surrounding the beam.

Dyer and Roeder considered the convenient scenario in which the
light beams travel through the homogeneous low density, or empty
regions, passing far away from the clumps, so that the shear becomes
negligible. However, we must consider whether the shear in our
particle simulation time-slices is able to significantly affect our
chosen values for the angular diameter distances.

Schneider and Weiss (1988a) performed Monte Carlo simulations to
determine the amplification of sources in a clumpy universe made up of
several lens-planes, each containing a random distribution of
point-like particles. They were able to show that the fraction of
`empty cones,' i.e., possible ray trajectories far from the clumps
with negligible shear, in a clumpy universe is small, so that in
general, the effects of shear must be taken into account in the
expression for the angular diameter distances. For rays weakly
affected by shear and with low amplifications, the linear terms in the
shear almost cancel, but higher order terms become more
important. However, the probability for rays being affected by shear
is dramatically lower in model universes with $\bar{\alpha}=0.8$
compared with universes with $\bar{\alpha}=0$. (We shall show shortly
that the values of $\bar{\alpha}$ in our sCDM simulations are always
at least 0.88, so that even at $z=0$ the matter distribution may be
considered smooth according to the usual definition of
$\bar{\alpha}$.) In summary, we might expect the number of rays
affected by shear to be low in smooth matter distributions, and then
the effect to be only of second order. Schneider and Weiss (1988a)
also derive an integral equation for the angular diameter distance
which they show to be equivalent to that of Dyer and Roeder (1973)
(without the shear term) when measured through the `empty cones.'

Watanabe and Tomita (1990) numerically solve the null geodesic
equations for light passing through a spatially flat Einstein-de
Sitter background universe in which the matter is condensed into
(softened) compact objects of galactic or galactic cluster dimensions,
and having an average specified separation in the present epoch. Their
conclusion, that, on average, the effect of shear on the
distance-redshift relation is small, providing the scale of the
inhomogeneities is greater than or equal to galactic scales, agrees
also with those of Futamase and Sasaki (1989), who show that, in most
cases, the shear does not contribute to the amplification. This
conclusion remains valid even when the density contrast is greater
than unity, although in the model used by Watanabe and Tomita (1990)
all amplifications were less than 2.

Our own work is conducted using a cosmological simulation in which the
distribution of matter is very smooth. Furthermore, our
minimum softening scale is of the order of galactic dimensions, so
that we feel justified in accepting that the shear plays only a second
order role in the distance-redshift relation in our sCDM data-set. (We
are able to quantify the effects of shear from our results in Section
6, and find that they are negligible.) With $\sigma \sim 0$,
therefore, equation (3) immediately reduces to the well-known
Dyer-Roeder equation. However, we also need to establish a value for
the smoothness parameter in our simulations, so that the appropriate
angular diameter distances can be evaluated and applied to the
data. Assuming $\sigma =0$, Schneider et al. (1992) give the following
generalised solution of the Dyer-Roeder equation for the angular
diameter distance between redshifts of $z_1$ and $z_2$ for $\Omega
=1$:
\begin{equation}
D(z_1,z_2)=\frac{c}{H_0}\frac{1}{2\beta}\left[\frac{(1+z_2)^{\beta
-\frac{5}{4}}}{(1+z_1)^{\beta +\frac{1}{4}}}-\frac{(1+z_1)^{\beta
-\frac{1}{4}}}{(1+z_2)^{\beta +\frac{5}{4}}}\right],
\end{equation}  
in which $\beta$ is expressed in terms of arbitrary $\bar{\alpha}$:
\begin{equation}
\beta =\frac{1}{4}(25-24\bar{\alpha})^{\frac{1}{2}}.
\end{equation}
We can write the left hand side of equation 4, equivalently, as
$D(z_1,z_2)=\frac{c}{H_0}r(z_1,z_2)$, in which $r(z_1,z_2)$ is the
dimensionless angular diameter distance. We show in Figure 1 the value
of the dimensionless multiplying factor, $R=r_dr_{ds}/r_s$, as it
applies to different time-slices at different redshifts, assuming
sources at $z_s=3.9$, 3.0, 1.9, 1.0 and 0.5. (These values correspond
to the redshifts of our time-slices, and have been chosen to be close 
to $z=4$, 3, 2, 1 and 0.5.) We have assumed zero shear, a completely
smooth distribution of matter, ($\bar{\alpha}=1$), and $\Omega =1$. 
We see that the peak in this factor occurs near $z=0.5$ for a source 
at redshift 4.

\begin{figure}
$$\vbox{ \psfig{figure=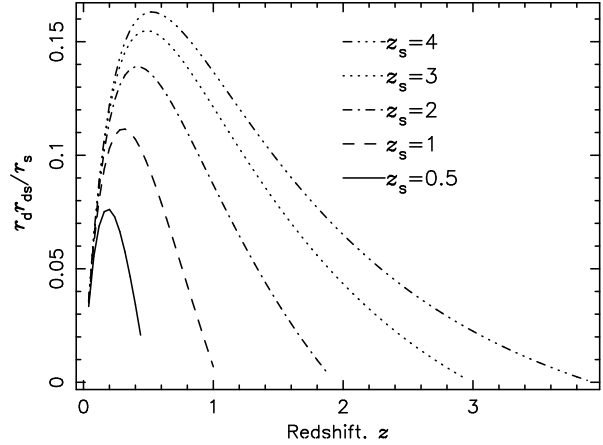,width=8.7truecm,angle=270} }$$
\caption{The dimensionless multiplying factor, $R=r_dr_{ds}/r_s$,
assuming sources at different redshifts. The uppermost curve for
a source at $z_s=4$, peaks at $z=0.52$; the next curve is for $z_s=3$
and peaks at $z=0.48$; the next curve is for $z_s=2$ and peaks at
$z=0.40$; the next curve is for $z_s=1$ and peaks at $z=0.32$; the
lowest curve is for $z_s=0.5$ and peaks at $z=0.20$.}
\label{fig:angdiamzs.qdp}
\end{figure}

From the output of our algorithm we are able to obtain an estimate of
the clumpiness or smoothness in each time-slice. Having set the
minimum softening scale, the code declares the number of particles
which are assigned the minimum softening, and we can therefore
immediately obtain the mass fraction contained in clumps, which we
choose to define by the minimum softening scale.

In the earliest time-slice at $z=3.6$, (next to $z=3.9$), there is a
mass fraction of only $5.6 \times 10^{-3}$ in clumps, giving
$\bar{\alpha}(z=3.6)=0.99$, and at $z=0$ the fraction is 0.12, giving
$\bar{\alpha}(z=0)=0.88$. Whilst we have not accurately tried to
assess the mean value for $\bar{\alpha}$ extending to different source
redshifts, it is clear that the value throughout is very close to 1,
and almost equivalent to the `filled beam' approximation. This result
concurs with Tomita (1998) who solves the null-geodesic equations for
a large number of pairs of light rays in four different cosmological
simulations with the sCDM spectrum. He uses $32^3$ particles in each,
softened to various physical radii up to a maximum of $40h^{-1}$kpc,
and finds $\bar{\alpha}$ to be close to 1 in all cases. However, there
does appear to be considerable dispersion in the values at late
times. We show in Figure 2 how similar the multiplying factor is for
the values $\bar{\alpha}=0.9$ and 1.0, and how these compare with a
value of $\bar{\alpha}=0$ for an entirely clumpy universe. The
discrepancy at the peak between $\bar{\alpha}=0.9$ and
$\bar{\alpha}=1.0$ is 3.1\%.

\begin{figure}
$$\vbox{
\psfig{figure=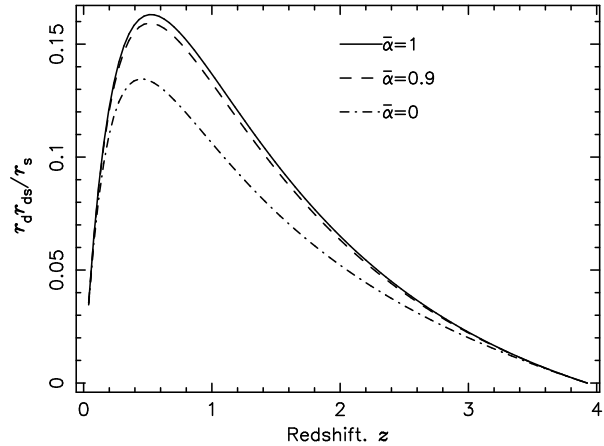,width=8.7truecm,angle=270}
}$$
\caption{The multiplying factor, $r_dr_{ds}/r_s$, for a source at
redshift 4, with smoothness parameters of 1 (uppermost curve), 0.9
(middle curve), and 0 (lowest curve).}
\label{fig:angdiamas4.qdp}
\end{figure}

Figure 3 shows the ratio of the multiplying factor for
$\bar{\alpha}=1$ and $\bar{\alpha}=0.9$ for the various source
redshifts. For sources at $z_s=2$ the maximum value of the ratio is
1.014, and for sources nearer than $z_s=1$ the discrepancy is well
below 1\%.

\begin{figure}
$$\vbox{ \psfig{figure=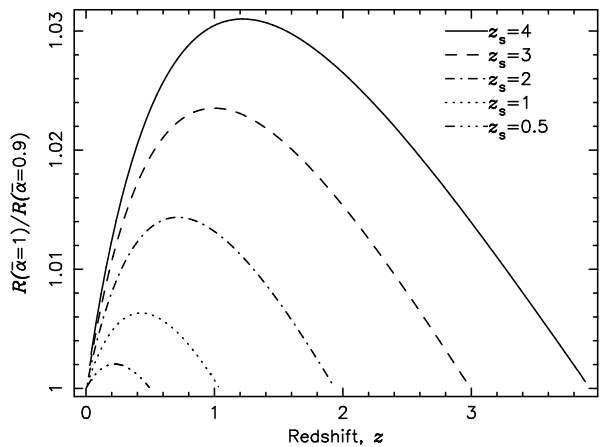,width=8.7truecm,angle=270} }$$
\caption{The ratio of the dimensionless angular diameter distance
multiplying factors, $R$, with $\bar{\alpha}=1$ and
$\bar{\alpha}=0.9$, for sources at redshifts 4, 3, 2, 1, and 0.5. The
values of the ratio at the peaks of these curves are 1.031, 1.024,
1.014, 1.006 and 1.002 respectively.}
\label{fig:angrat4.qdp}
\end{figure}

\section{THE FORMATION OF STRUCTURE}
    
The shear algorithm generates the six independent three-dimensional
shear component values (expressed in box units), and we have chosen to
compute them at 1000 evaluation positions along $100 \times 100$ lines
of sight in each simulation time-slice. In a simplistic way, the
magnitude of these components characterises the particular
time-slice. To convert the components to absolute values we have to
apply the appropriate angular diameter distance factors,
$R=r_dr_{ds}/r_s$, as described in the previous section, together with
the factor $B(1+z)^2$, where $B=3.733 \times 10^{-9}$ for the
simulation boxes we have used (which have comoving dimensions of
$100h^{-1}$Mpc) and where the $(1+z)^2$ factor occurs to convert the
comoving code units to physical units.

The magnitude of the rms value determined from each component multiplied by
$B(1+z)^2$ in each time-slice is then of interest. In Figure 4 we
show these values for the sum of the diagonal terms in the (projected)
matrix of effective lensing potentials; this is closely associated
with the surface density, which in turn determines the magnifications
produced in the time-slice. We notice that the values for these
combined components very slowly decreases towards $z=0$. This same
trend is apparent with the other components individually. It has the
interesting interpretation that, even though structure is forming (to
produce greater magnification locally), the real expansion of the
universe (causing the mean particle separation to increase) just
outweighs this in terms of the magnitudes of the component
values. Nevertheless, the formation of structure can be seen; by
considering just the sets of highest values in each time-slice, again
multiplied by the factor $B(1+z)^2$, and taking the mean values of
these, we see in Figure 4 an initial fall as the universe expands and
before structure has begun to form, and then at later times an
increase in the mean values, indicative of the existence of dense
(bound) structures.

\begin{figure}
$$\vbox{
\psfig{figure=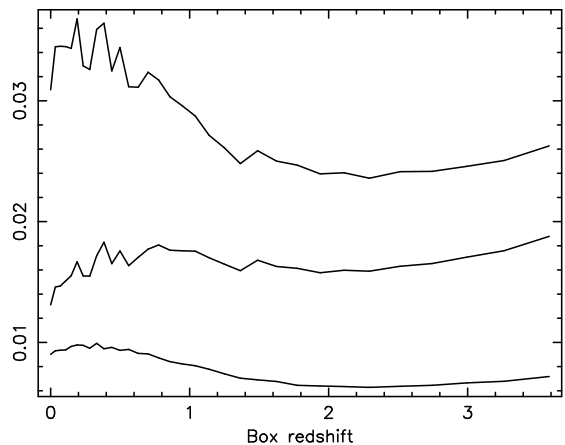,width=8.7truecm,angle=270}
}$$
\caption{Middle curve: the rms value in each time-slice, multiplied by
$B(1+z)^2$, of the sum of the diagonal components of the
two-dimensional shear components, showing a gradual fall towards
$z=0$. Top curve: The set of highest values of the summed diagonal
components, multiplied by $B(1+z)^2$, which shows the initial gradual
fall as the universe expands, and then an increase towards $z=0$ as
structure forms. Lowest curve: The set of highest values for one of
the off-diagonal components, multiplied by $B(1+z)^2$, which shows a
similar trend to the top curve.}
\label{fig:high2.qdp}
\end{figure}

However, when the values are then multiplied by the angular diameter
distance factor, $R$, we see in Figure 5 that the peaks are
extremely broad, indicating that significant contributions to the
magnifications and ellipticities can arise in time-slices covering a
wide range of redshifts, and not just near $z=0.5$ where $R$ has its
peak (for sources at $z_s=4$).

\begin{figure}
$$\vbox{
\psfig{figure=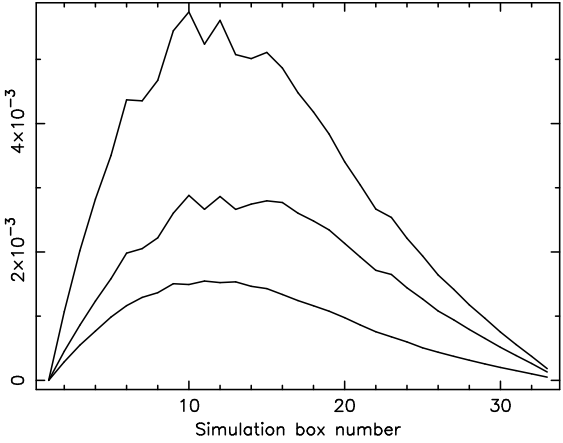,width=8.7truecm,angle=270}
}$$
\caption{Components in each time-slice converted to absolute values,
including the angular diameter distance factors, for sources at
$z_s=4$. Middle curve: the rms value for each time-slice of the sum of
the diagonal components of the two-dimensional shear components. Top
curve: The set of highest values of the summed diagonal
components. Lowest curve: The set of highest values for one of the
off-diagonal components. All the curves are seen to exhibit extremely
broad peaks, indicative of significant contributions from a wide range
of redshifts. (Box 10 is at redshift 0.3835, and box 20 is at redshift
1.1400.)}
\label{fig:high3.qdp}
\end{figure}

Premadi, Martel and Matzner (1998a) have done similar work, using 5
different sets of initial conditions for each of their $N$-body
simulations, so that the time-slices can be chosen at random from any
one of the 5 sets, randomly translated to avoid correlations in the
large scale structure between adjacent boxes, and projected onto
planes. They solve the two-dimensional Poisson equation on a grid, and
use a FFT method to obtain the first and second derivatives of the
gravitational potential on each plane. They consider the effects on
light beams, each consisting of 65 rays arranged in concentric rings
to represent circular images, and have performed 500 calculations for
each cosmological model, based on 500 different random translations of
the planes. For the shear and magnification they find that the
individual contribution due to each lens-plane is greatest at
intermediate redshifts, of order $z=1-2$, for sources located at
$z_s=5$.

Premadi, Martel and Matzner (1998b, c) also report their results for
the shear for sources at $z_s=3$, and again find very broad peaks
covering a wide range of (intermediate) lens-plane redshifts.

\section{MULTIPLE LENS-PLANE THEORY}

As described in Section 2.2, we establish 1000 evaluation positions
along each of the $100 \times 100$ lines of sight through each
simulation time-slice, and the shear algorithm computes the six
independent second derivatives of the gravitational potential at each
position. By integration of the values we establish the matrix of
two-dimensional effective lensing potentials at each of 50 positions
along every line of sight. We establish the Jacobian matrix, $\cal A$,
from these effective lensing potentials by applying the appropriate
multiplying factors, as described in Section 3, and the Jacobian
develops along the line of sight for each evaluation position. It is
computed recursively in accordance with the multiple lens-plane
theory, which is summarised by Schneider et al. (1992). The final
Jacobian matrix after $N$ deflections is
\begin{equation}
{\cal A} _{\rm total} = {\cal I} - \sum_{i=1}^{N}{\cal U} ^i {\cal
A}_i,
\end{equation}
where $\cal I$ is the unit matrix,
\begin{equation}
{\cal U}^i = \left( \begin{array}{cc}
	\psi_{11}^i  & \psi_{12}^i \\
	\psi_{21}^i  & \psi_{22}^i
	\end{array}
	\right)
\end{equation}
for the $i$th deflection, and the
intermediate Jacobian matrices are
\begin{equation}
{\cal A}_j = {\cal I} - \sum_{i=1}^{j-1}\beta_{ij}{\cal U}_i{\cal
A}_i,
\end{equation}
where
\begin{equation}
\beta  _{ij} = \frac{D_s}{D_{is}}\frac{D_{ij}}{D_j},
\end{equation}
in which $D_j$, $D_{is}$ and $D_{ij}$ are the angular diameter
distances to the $j$th lens, that between the $i$th lens and the
source, and that between the $i$th and $j$th lenses, respectively.

The magnification, $\mu$, at any position, is given in terms of the
Jacobian at that point:
\begin{equation}
\mu =\left(\det \cal A \right)^{-1},
\end{equation}
so that we can assess the magnification as it
develops along a line of sight, finally computing the emergent
magnification after passage through an entire box or set of boxes.
The convergence, $\kappa$, is defined by
\begin{equation}
\kappa = \frac{1}{2}(\psi_{11}+\psi_{22})
\end{equation}
from the diagonal elements of the Jacobian matrix, and causes
isotropic focussing of light rays, and so isotropic magnification of
the source. Thus, with convergence acting alone, the image would be
the same shape as, but a different size from, the source.

The shear, $\gamma$, in each line of sight, is given by
\begin{equation}
\gamma^2 = \gamma_1^2 + \gamma_2^2 \equiv
\frac{1}{4}(\psi_{11}-\psi_{22})^2 + \psi_{12}^2.
\end{equation}
Shear introduces anisotropy, causing the image to be a different
shape, in general, from the source.

From equation 10, and these definitions, 
\begin{equation}
\mu = (1-\psi_{11}-\psi_{22}+\psi_{11}\psi_{22}-\psi_{12}^2)^{-1},
\end{equation}
so that with weak lensing the magnification reduces to
\begin{equation}
\mu \simeq 1+2\kappa +3\kappa^2 + \gamma^2 + O(\kappa^3,\gamma^3).
\end{equation}
In the presence of convergence and shear, a circular source becomes
elliptical in shape, and the ellipticity, $\epsilon$, defined in terms
of the ratio of the minor and major axes, becomes
\begin{equation}
\epsilon = 1 - \frac{1-\kappa -\gamma}{1-\kappa +\gamma},
\end{equation}
which reduces to
\begin{equation}
\epsilon \simeq 2 \gamma (1+\kappa -\gamma ) + O(\kappa^3,\gamma^3)
\end{equation}
in weak lensing.

The multiple lens-plane procedure allows values and distributions of
the magnification, ellipticity, convergence and shear to be obtained
at $z=0$ for light rays traversing the set of linked simulation boxes
starting from the chosen source redshift. The ability to apply the
appropriate angular diameter distances at every evaluation position
avoids the introduction of errors associated with planar methods, and
also allows the possibility of choosing source positions within a
simulation box if necessary. This may be useful when considering the
effects of large-scale structure on real observed sources at specific
redshifts, or if the algorithm is to be applied to large simulation
volumes.
  
\section{RESULTS}

We first examine the importance of the smoothness parameter,
$\bar{\alpha}$, in the distance-redshift relation, to the
magnification distribution, by computing the magnifications due to a
single (assumed isolated) simulation box at $z=0.5$ for a source at
$z_s=4.$ (At this box redshift the contribution to the magnifications
is expected to be near the maximum.) The magnification distributions
arising for $\bar{\alpha}=1$ and $\bar{\alpha}=0.9$, (deduced from our
simulations, as explained in Section 3) are virtually
indistinguishable. The only significant difference is the maximum
value of the magnification in each case, which is only 1.9\% higher in
the $\bar{\alpha}=1$ case. We therefore feel justified in presenting
our results based on a smoothness parameter of $\bar{\alpha}=1$
throughout.

We have chosen to assume source redshifts, $z_s$, close to 4, 3, 2, 1
and 0.5, and shall refer to the sources in these terms. The actual
redshift values are 3.9, 3.0, 1.9, 1.0 and 0.5 respectively,
corresponding to nominal time-slice redshifts in our sCDM
simulation. For each source redshift we have evaluated the final
emergent Jacobian matrix at $z=0$ for all 10000 lines of sight, by
linking all the simulation boxes between the source redshift and
$z=0$, as described in Section 5, and, by manipulation of the data
according to the multiple lens-plane equations, we have been able to
produce all the required values for the magnifications, ellipticities,
shear and convergence.

Figures 6 and 7 show the distributions of the magnifications, $\mu$,
for the five source redshifts, and for all source redshifts there is a
significant range. The rms fluctuations for the magnifications about
the mean value of $<\mu>=1$ are displayed in column 2 of the table for
each source redshift. However, since the magnification distributions
are asymmetrical, we have calculated the values, $\mu_{\mathrm{low}}$
and $\mu_{\mathrm{high}}$, above and below which
97\mbox{$\frac{1}{2}$}\% of all lines of sight fall. These are
displayed in columns 3 and 4 of the table.

\begin{figure}
$$\vbox{
\psfig{figure=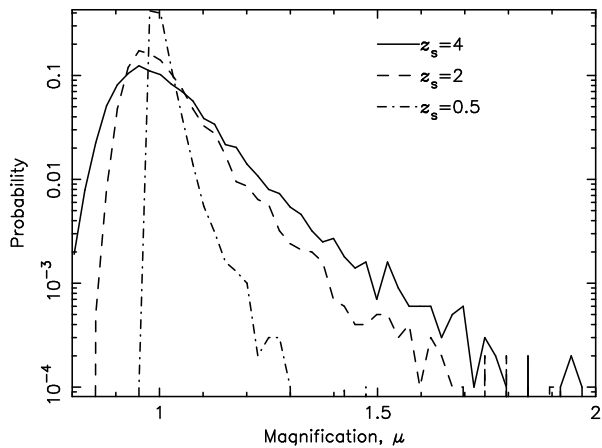,width=8.7truecm,angle=270}
}$$
\caption{Probability distributions for the magnification, for $z_s=4,$
2 and 0.5.}
\label{fig:magdistz4.qdp}
\end{figure}

\begin{figure}
$$\vbox{
\psfig{figure=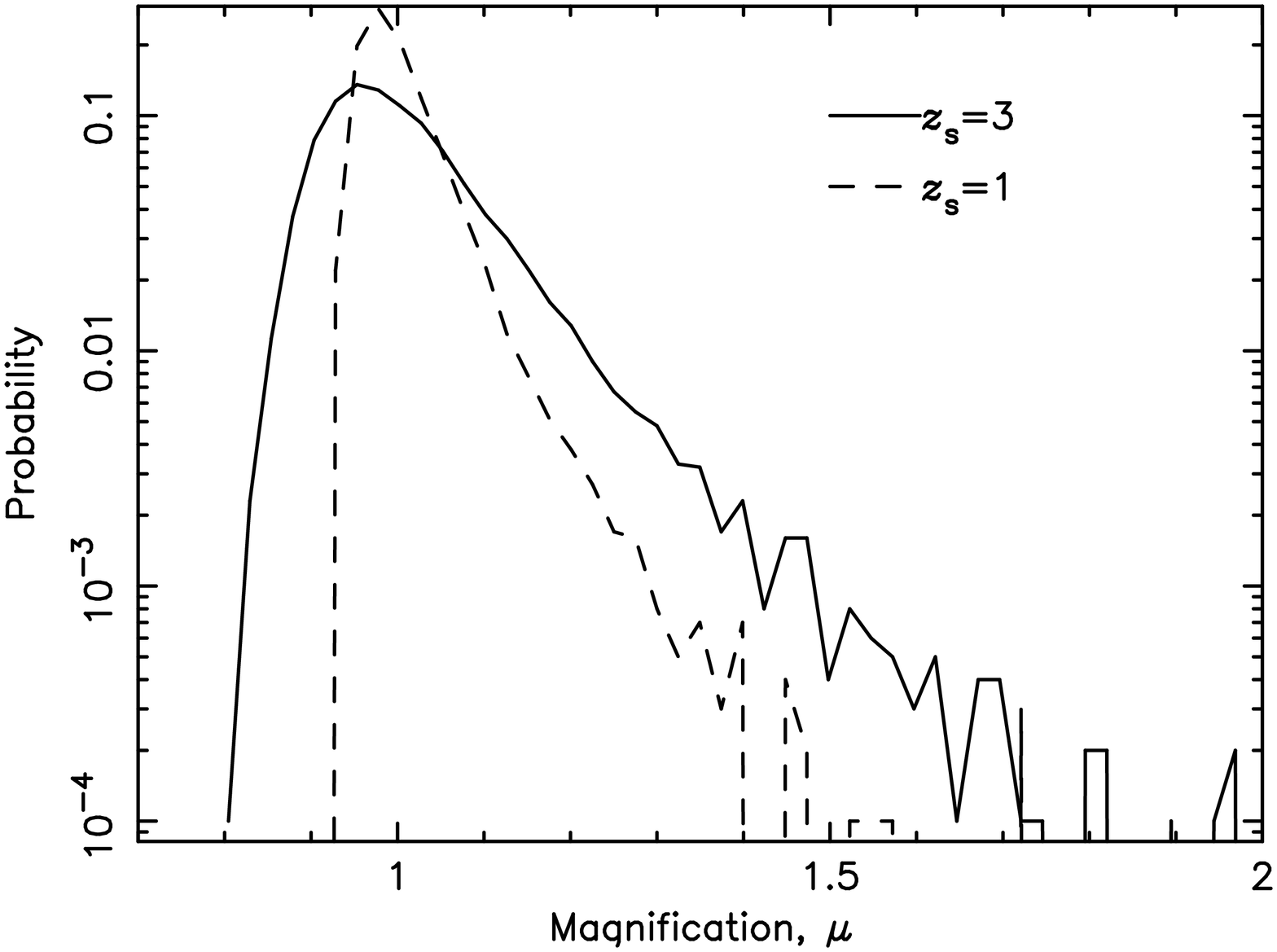,width=8.7truecm,angle=270}
}$$
\caption{Probability distributions for the magnification, for $z_s=3,$
and 1.}
\label{fig:magdistz4.qdp}
\end{figure}

\begin{table}
\begin{center}
\begin{tabular}{|l|c|c|c|}
\hline
$z_s$ & $\mu_{\mathrm{rms}}$ & $\mu_{\mathrm{low}}$ &
$\mu_{\mathrm{high}}$ \\ \hline
4   & 0.126 & 0.85 & 1.30 \\
3   & 0.111 & 0.86 & 1.26 \\
2   & 0.088 & 0.88 & 1.20 \\
1   & 0.056 & 0.93 & 1.13 \\
0.5 & 0.027 & 0.97 & 1.05 \\
\hline
\end{tabular}
\end{center}
\caption{Column 2 shows the rms fluctuations for the
magnifications, $\mu_{\mathrm{rms}}$, about the mean value of
$<\mu>=1$ for the various source redshifts, $z_s$; column 3 shows the
magnification value $\mu_{\mathrm{low}}$ for each redshift above which
97\mbox{$\frac{1}{2}$}\% of all lines of sight fall; column 4 shows
the magnification values $\mu_{\mathrm{high}}$ for each redshift below
which 97\mbox{$\frac{1}{2}$}\% of all lines of sight fall;}
\end{table}

The accumulating number of lines of sight having magnifications
greater than the abscissa value is shown in Figure 8 for the five
different source redshifts, and clearly shows the distinctions at the
high magnification end.

\begin{figure}
$$\vbox{
\psfig{figure=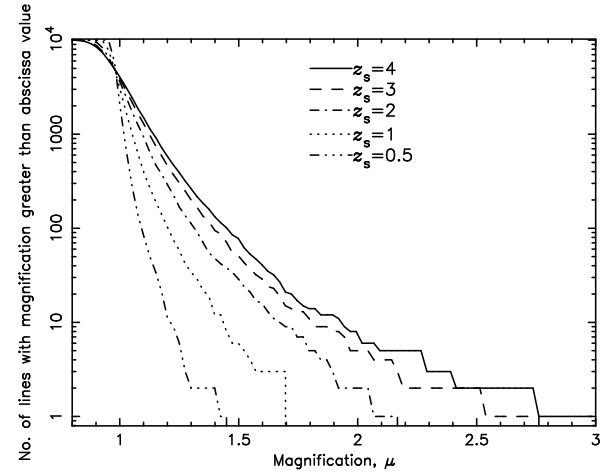,width=8.7truecm,angle=270}
}$$
\caption{The accumulating number of lines of sight with magnifications
greater than the abscissa value, for $z_s=4,$ 3, 2, 1 and 0.5.}
\label{fig:magdistz4.qdp}
\end{figure}

In Figure 9 we show the magnification, $\mu$, plotted against the
convergence, $\kappa$, for $z_s=4$, and see that the magnification is
clearly not linear in $\kappa$ as expected for small magnitudes of
$\kappa$. This is true for all our source redshifts except $z_s=0.5$,
for which the curve is closely linear throughout. The non-linearity
arises because of the presence of the higher order terms in the
expression for $\mu$ given by equation 14, and we show for comparison
the curve of $\mu = 1+2\kappa +3\kappa^2$.

\begin{figure}
$$\vbox{
\psfig{figure=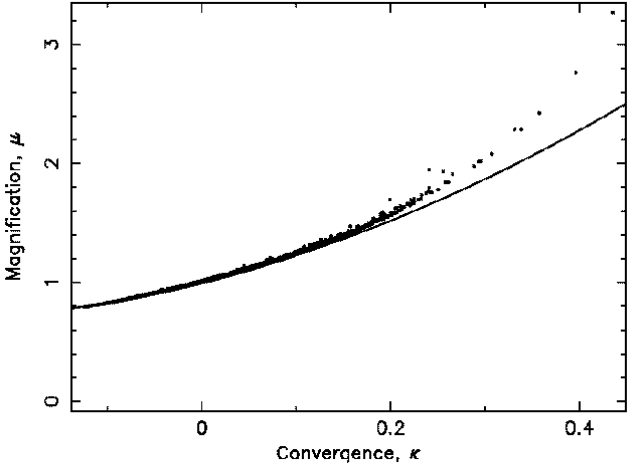,width=8.7truecm,angle=270}
}$$
\caption{$\mu$ vs. $\kappa$ for $z_s=4$ (dots). The continuous line,
shown for comparison, represents $\mu = 1 + 2\kappa + 3\kappa^2$.}
\label{fig:magz4.qdp}
\end{figure}

We would generally expect the shear, $\gamma$, to fluctuate strongly
for light rays passing through regions of high density (high
convergence), and we indeed find considerable scatter in the shear
when plotted against the convergence. Figure 10, however, shows the
result of binning the convergence values and calculating the average
shear in each bin, for sources at $z_s=4$. We see that throughout most
of the range in $\kappa$ the average shear increases very slowly, and
closely linearly. (At the high $\kappa$ end there are too few data
points to establish accurate average values for $\gamma$.) This result
suggests that there may be a contribution to the magnification from
the shear, and we discuss this later in this Section.

\begin{figure}
$$\vbox{
\psfig{figure=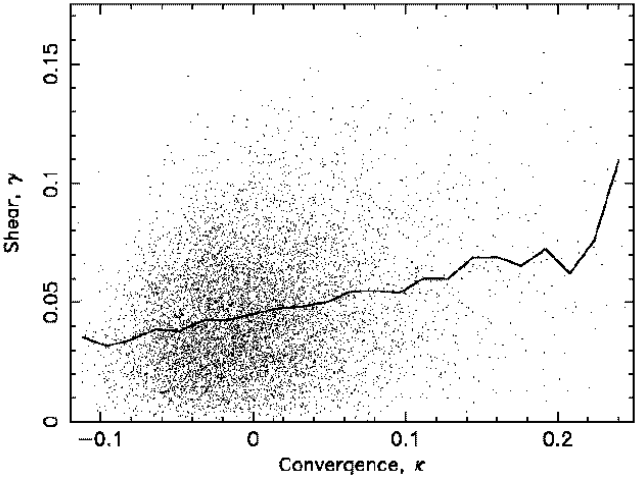,width=8.7truecm,angle=270}
}$$
\caption{Shear vs. convergence for sources at $z_s=4$ (dots), and the
average shear (full line) in each of the $\kappa$ bins, which shows a
slow and nearly linear increase with increasing convergence.}
\label{fig:onesiggam.qdp}
\end{figure}

Figure 11 shows the distributions in the convergence, $\kappa$,
primarily responsible for the magnifications. The rms values for the
convergence are 0.052 (for $z_s=4$), 0.047 (for $z_s=3$), 0.038 (for
$z_s=2$), 0.025 (for $z_s=1$) and 0.013 (for $z_s=0.5$). These values
are entirely consistent with the rms fluctuations for the
magnification about the mean (stated above), being slightly below half
the rms magnification values (see equation 14).

\begin{figure}
$$\vbox{
\psfig{figure=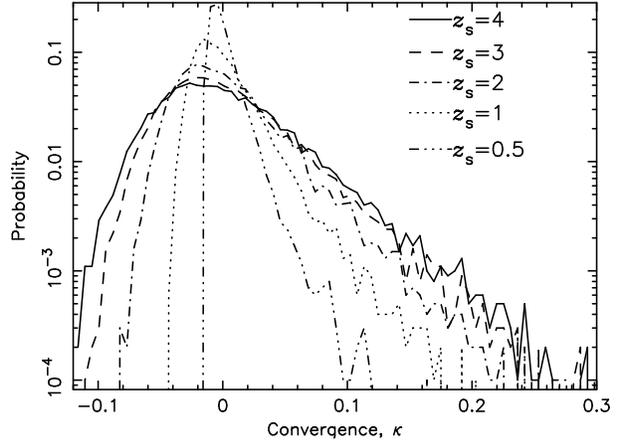,width=8.7truecm,angle=270}
}$$
\caption{The probability distributions for the convergence, $\kappa$,
for the five different source redshifts.}
\label{fig:kappadistz4.qdp}
\end{figure}

The distributions in the shear, $\gamma$, (defined according to
equation 12) for the five source redshifts, are broadest, as expected,
for the highest source redshifts, and, for $z_s=4$,
97\mbox{$\frac{1}{2}$}\% of all lines of sight have shear values below
0.103. The ellipticity, $\epsilon$, in the image of a source is
primarily produced by the shear, and we show in Figure 12 the
distributions in $\epsilon$ for the five source redshifts. The peaks
in the ellipticity distributions occur at $\epsilon =0.057$ for
$z_s=4$, 0.057 for $z_s=3$, 0.047 for $z_s=2$, 0.027 for $z_s=1$ and
0.012 for $z_s=0.5$. Figure 13 displays the accumulating number of
lines of sight with $\epsilon$ greater than the abscissa value. For 
$z_s=4$, we find that 97\mbox{$\frac{1}{2}$}\% of all lines
of sight have ellipticities up to 0.195. In Figure 14 we see that the
ellipticity is very closely linear in terms of $\gamma$ throughout
most of the range in $\gamma$. The scatter arises because of the
factor containing the convergence, $\kappa$, in equation 16.

\begin{figure}
$$\vbox{
\psfig{figure=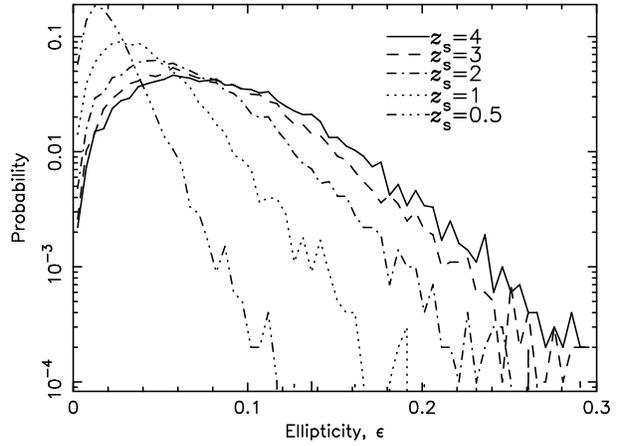,width=8.7truecm,angle=270}
}$$
\caption{The probability distributions for the ellipticity,
$\epsilon$, for the five different source redshifts.}
\label{fig:epsilondistz4.qdp}
\end{figure}

\begin{figure}
$$\vbox{
\psfig{figure=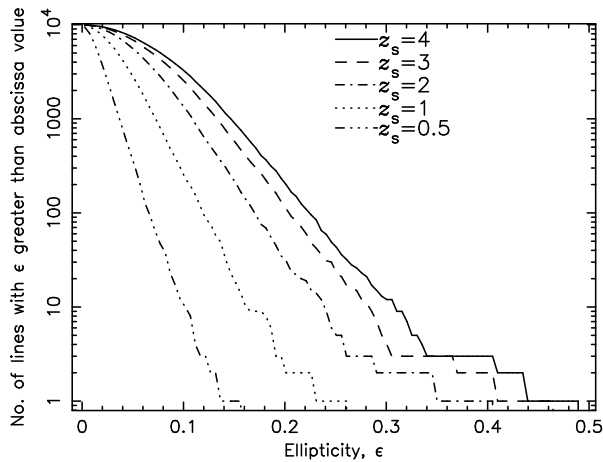,width=8.7truecm,angle=270}
}$$
\caption{The accumulating number of lines of sight with $\epsilon$
greater than the abscissa value, for the five source redshifts.}
\label{fig:epsilondistz4.qdp}
\end{figure}

\begin{figure}
$$\vbox{
\psfig{figure=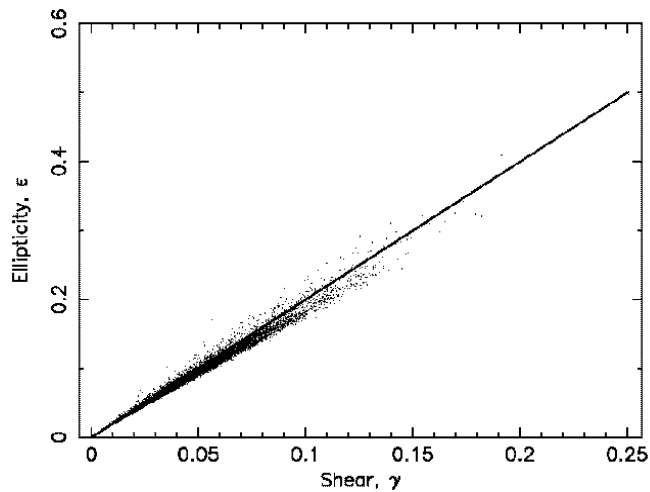,width=8.7truecm,angle=270}
}$$
\caption{Source ellipticity vs. shear for $z_s=4$ (dots). The straight
line, shown for comparison, represents $\epsilon = 2\gamma.$}
\label{fig:magz4.qdp}
\end{figure}

Finally, we attempted to see if there was a contribution to the
magnification from the shear as implied by the distance-redshift
relation (equation 3). We found considerable scatter, as expected, in
the plots of magnification vs. shear, but we found in Figure 10 a
tenuous connection between the shear and the convergence, indicating
that there may be a similar connection between the magnification and
the shear. We see from equation 14 that the effect of shear is only of
second order (as established by Schneider and Weiss, 1988a). By
binning the shear values and calculating the average magnification in
each bin, we are able to show (Figure 15) that there may be a slow
increase in $<\mu>$ with increasing shear. Figure 15 is for sources at
$z_s=4$. Although there are insufficient data points at the high shear
end, it still seems likely that the effects of shear on the mean
magnification may be at least 10\% for shear values greater than about
0.1. However, interestingly, only 2.6\% of the data points in our
simulation produced shear in excess of 0.1. According to equation 3,
the shear has an effect in the distance-redshift relation equivalent
to increasing the effective smoothness parameter,
$\bar\alpha$. However, by substituting the mean shear value determined
for sources at $z_s=0.5$ the effect on $\bar\alpha$ is found to be
completely negligible. Furthermore, the importance of the effect
reduces with redshift, so that our conclusion in Section 3, to ignore
the effects of shear in the distance-redshift relation, can now be
justified.

\begin{figure}
$$\vbox{
\psfig{figure=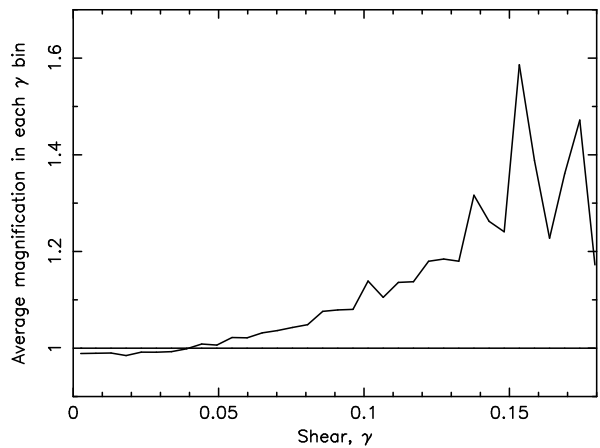,width=8.7truecm,angle=270}
}$$
\caption{The average magnification in each shear bin, for sources at
$z_s=4$. The overall mean magnification at $<\mu>=1$ is shown for
comparison.} 
\label{fig:onesigmag.qdp}
\end{figure}


\section{DISCUSSION OF RESULTS}

Following the brief summary in the Introduction of work by other
authors on the effects of weak gravitational lensing in the sCDM
cosmology, we described in Section 2 the algorithm for the
three-dimensional shear, and some of the key advantages it offers over
other methods. In particular, we mentioned the ability of the code to
include automatically the effects of matter in the periodic images of
the fundamental volume, so that matter effectively stretching to
infinity is included in computations of the shear. We also described
the variable softening feature in the code which allows a good
physical interpretation of the matter distribution in simulation
time-slices to be made. We explained our choice of an appropriate
minimum for the variable softening, taking into account its physical
dimension, the degree of particle clustering, and the likelihood of
strong lensing effects. We also described in Section 2 the sCDM
simulations we have used, which are available from the Hydra 
consortium.

One clear advantage of the algorithm operating on a three-dimensional
volume is that we can apply the appropriate angular diameter distance
to every single evaluation position, thereby avoiding the introduction
of errors associated with the use of single values in two-dimensional
methods. (Couchman et al., 1998, analyses these possible
errors.) However, we have had to consider what the `appropriate' values
should be. 

First, we considered the effects of shear in the distance-redshift
relation (equation 3), and were guided by the findings of Schneider
and Weiss (1988a), Watanabe and Tomita (1990) and Futamase and Sasaki
(1989) that the shear probably has only a second order effect. Our
decision to ignore the effects of shear in the relation in general is
justified because we found that significant effects may occur only in
$\sim 2.6\%$ of the lines of sight, and the impact on the effective
value of the smoothness parameter, $\bar\alpha$, by substituting the
mean values of the shear, is completely negligible at all redshifts.

Second, we needed to include a suitable value for the smoothness
parameter, $\bar{\alpha}$. The minimum value for the variable
softening in the shear algorithm, and the number of particles falling
within this minimum value, provides an excellent framework for
determining $\bar{\alpha}$ in accordance with the original definition
of Dyer and Roeder (1972).

We find, on this definition, that $\bar{\alpha}$ varies between
approximately 1.0, in the $z=3.6$ time-slice, and 0.9, at $z=0$, and
we therefore checked the significance for the angular diameter
distance multiplying factor with these extreme values. For sources at
$z_s=4$ the difference between the factors is very small (see Figure
2) at all lens redshifts, and the maximum discrepancy is only 3.1\%,
as shown in Figure 3. This discrepancy is always less that 1\% for
sources with redshifts less than 1. Furthermore, we investigated the
effects of $\bar{\alpha}$ on the magnification distribution arising
from a single (assumed isolated) simulation box at $z=0.5$ for a
source at $z_s=4$. The two distributions are virtually
indistinguishable; the most significant difference is in the values of
the maximum magnification in each case, which differs by only
1.9\%. For these reasons we chose to proceed with our analysis of the
results for the sCDM cosmology on the basis of $\bar{\alpha}=1$.

In Section 4 we found the general and interesting result that the rms
values of the `intrinsic' computed shear values, multiplied by the
conversion factor $B(1+z)^2$, but before the application of the
angular diameter distance multiplying factors, fell slowly with
redshift towards $z=0$, (i.e., with the evolving and expanding
universe). Evidently, the universal expansion just outweighs the
formation of structure when viewed in terms of the shearing on
light. The formation of structure could be seen by considering only
the sets of highest values in each time-slice, and then the mean
values of these initially fall, before increasing slowly at the onset
of structure formation. When the appropriate angular diameter distance
multiplying factors were applied to the computed values, we then found
the further interesting result that there can be considerable
contributions to the shear and magnification arising from time-slices
covering a very broad range of redshifts. This result is displayed in
Figure 5.

In Section 5 we described how the data computed from our sets of
simulation boxes were manipulated in accordance with the multiple
lens-plane theory to produce the results of Section 6. There we showed
results based on sources at five different redshifts, namely $z_s=4,$
3, 2, 1 and 0.5. We showed distributions in the magnification (and
details of the high magnification end of these distributions), the
convergence and the ellipticity (which closely resembles the
distribution in the shear), and also the relationships amongst these
various quantities. Figure 9 shows the strong departure from the
linear regime for the magnification as a function of the convergence,
whilst Figure 14 shows a closely linear relationship between the
ellipticity and the shear. Figure 10 suggests a slow increase in shear
with increasing convergence, broadly as expected. For sources at
$z_s=4$, 97\mbox{$\frac{1}{2}$}\% of all lines of sight have
magnification values up to 1.30. (The maximum magnifications depend on
the choice of the minimum softening in the code, although the overall
distributions are very insensitive to the softening.) In particular,
we found rms fluctuations in the magnification (about the mean) as
much as 0.13 for sources at $z_s=4$.  Even for sources at $z_s=0.5$
there is a measurable range of magnifications up to 1.05 for
97\mbox{$\frac{1}{2}$}\% of the lines of sight.

We summarised in the Introduction the methods of other workers using
the sCDM cosmology. Because of the way in which Jaroszy\'nski et
al. (1990) determine the magnifications, their distributions do not
have mean magnifications of 1. However, their dispersions in the
convergence for sources at $z_s=1$ and $z_s=3$ can be seen to be
considerably lower than our values. In addition the dispersions appear
to show very little evolution with redshift. Wambsganss et al. (1998)
find magnifications up to 100 and correspondingly highly dispersed
distributions, very much larger than ours at $z_s=3$. (Their
magnification distributions show separately the results for
multiply-imaged sources and singly-imaged sources.) The very wide
distributions they find have also enabled them to support a $\mu^{-2}$
power-law tail in the distribution which is predicted by Schneider et
al. (1992) in the case of magnification by point sources when $\mu \gg
1.$ The magnification distributions of Premadi et al. (1998a) appear
incomplete, but the range in magnifications appears to be rather
similar to ours for sources at $z_s=3.$ This is reassuring because,
although their method relies on two-dimensional projections of the
simulation boxes, they include many of the essential features to which
we have drawn attention, for example, an assumed periodicity in the
matter distribution, randomly chosen initial conditions to avoid
structure correlations between adjacent simulation boxes, the net zero
mean density requirement, realistic mass profiles for the particles,
and use of the filled beam approximation with a smoothness parameter,
$\bar\alpha =1$. Marri and Ferrara (1998) show very much wider
magnification distributions than we have found, and also very high
maximum values, which occur as a result of using point particles
rather than smoothed particles. We also disagree with their choice of
$\bar\alpha = 0$, which is representative of an entirely clumpy
universe, as opposed to our finding that the sCDM universe is very
close to being smooth (with $\bar\alpha \simeq 1$) at all epochs.

In our own work 97\mbox{$\frac{1}{2}$}\% of the lines of sight have
ellipticities up to 0.195 for $z_s=4$. At the peaks of the
distributions we found values of 0.057 and 0.027 for $\epsilon$ for
sources at $z_s =3$ and 1 respectively. These are somewhat lower than
the values of 0.095 ($z_s=3$) and 0.045 ($z_s=1$) found by
Jaroszy\'nski et al. (1990). Rather surprisingly, however, their peak
values in the distributions for the shear are quite similar to our
own, especially for sources at $z_s=3$.

Our magnification results may have an impact on the interpretation of
the magnitude data for high-redshift Type Ia Supernov\ae~reported by
Riess, Filippenko, Challis et al. (1998), since we have seen in
Section 6 the possible range of magnifications that may apply to
distant sources. The high-redshift Supernov\ae~data include sources up
to redshifts of 0.97, so that the effects of the large-scale structure
should not be ignored when interpreting the peak magnitudes and
distance moduli. However, our magnification values for $z_s=1$ and
$z_s=0.5$ above and below which 97\mbox{$\frac{1}{2}$}\% of all lines
of sight fall are considerably closer to unity than the values found
by Wambsganss, Cen, Xu and Ostriker (1997) for the sCDM model. We would therefore
expect to find correspondingly smaller lensing-induced dispersions in
the distance moduli. However, we hope to quantify the dispersions in
the distance moduli and the effect on the deceleration parameter,
$q_0$, for an open cosmology in a future paper, especially in view of
Riess et al.'s (1998) conclusions in favour of an open universe with a
cosmological constant.

Another area affected by the presence of a distribution in
magnifications is the luminosity function for quasars or high-redshift
galaxies. Most sources are demagnified (the median value for $\mu$ is
always just less than 1) which will remove many galaxies from the dim
end of the luminosity function in a flux-limited survey, but at
$z_s=2$, say, we find an rms fluctuation in the magnifications of
8.8\% which will also allow some dim galaxies to be magnified and
observed, where otherwise they would not have been.

In addition to considering these matters further we hope to address
the following questions in the immediate future.

1. How does the redshift dependence of the shear matrix change in
low-density universes? We shall be attempting to answer this question
using simulation data from other cosmologies available from the Hydra
consortium. In particular, we shall work on open and flat cosmological
simulations with $\Omega_0 = 0.3$. Of particular interest is the flat
model with $\Omega_0 = 0.3$ and cosmological constant $\Lambda_0 =
0.7$, in view of the recent work by Riess et al. (1998) indicating the
likelihood of this type of universe. In critical density universes it
is believed that clustering continues to grow to the present day, and
this is indicated by the results shown in Figure 4. However, in low
density universes, structures should have formed by $z \sim
\Omega_0^{-1} - 1$, so that the shapes of the curves in Figure 4 are
likely to be very different.

2. How do our distributions in the magnification, ellipticity, shear
and convergence vary amongst different cosmologies? With low-density
universes, weak lensing effects are likely to be very different due to
four main factors: (i) the formation of structure at earlier times,
and its persistence through periods in which the contribution to the
lensing is significant; (ii) dilution of the effects as the universe
expands beyond the formation of structure; (iii) different values for
the angular diameter distances; (iv) the lower average values for the
computed shear components in view of the lower density values in the
universe.

3. Do the high-magnification and low-ellipticity lines of sight occur
because of the effects of individual large clusters, or as a result of
continuous high density regions such as filamentary structures?

4. How frequently do lines of sight in the direction of
multiply-imaged quasars coincide with lines of high convergence
associated with the general form of the large-scale structure
(independent of the lensing galaxy)? There is clear evidence (Thomas,
Webster \& Drinkwater, 1995) of increased numbers of near-neighbour
galaxies (when viewed along the line of sight) to bright quasars, and
this raises the intriguing possibility that some sub-critical lenses
may become critical (and produce multiple images of background
sources) in the presence of high density large-scale structure along
the line of sight. According to the multiple lens-plane theory it is
entirely consistent that the determinant of the developing Jacobian
matrix along a high-convergence line of sight may change sign in the
presence of a high surface density (but sub-critical) lens. In such a
scenario modifications to the models for the surface density profile
of the lensing galaxy would also be required.

\section*{ACKNOWLEDGMENTS}

We are indebted to the Starlink minor node at the University of Sussex
for the preparation of this paper, and to the University of Sussex for
the sponsorship of AJB.  We thank NATO for the award of a
Collaborative Research Grant (CRG 970081) which has greatly
facilitated our interaction.  R. L. Webster and C. J. Fluke of the
University of Melbourne, and K. Subramanian of the National Centre for
Radio Astrophysics, Pune, have been particularly helpful.

\section*{REFERENCES}

\paper{Blandford R. D. \& Narayan R.}{1986}{Ap. J.}{310}{568}
\paper{Blandford R. D. \& Kochanek C. S.}{1987}{Proc. 4th Jerusalem
Winter School for Th. Physics, Dark Matter in the Universe,
ed. Bahcall J. N., Piran T. \& Weinberg S.}{Singapore, World
Scientific}{p.133} 
\paper{Couchman H. M. P., Barber A. J. \& Thomas P. A.}{1998}
{astro-ph}{9810063}{Preprint}
\paper{Couchman H. M. P., Thomas, P. A., \& Pearce F. R.}{1995}
{Ap. J.}{452}{797}
\paper{Dyer C. C. \& Roeder R. C.}{1972}{Ap. J. (Letts.)}{174}{L115}
\paper{Dyer C. C. \& Roeder R. C.}{1973}{Ap. J. (Letts.)}{180}{L31}
\paper{Falco E. E., Govenstein M. V. \& Shapiro
I. I.}{1991}{Ap. J.}{372}{364}
\paper{Fluke C. J., Webster R. L. \& Mortlock
D. J.}{1998a}{astro-ph}{9812300}{Preprint}
\paper{Fluke C. J.}{Webster R. L.}{Mortlock D. J.}{1998b}{In preparation}
\paper{Futumase T. \& Sasaki M.}{1989}{Phys. Rev. D}{40}{2502}
\paper{Grogan N. A. \& Narayan R.}{1996}{Ap. J.}{464}{92}
\paper{Hockney R. W. \& Eastwood J. W.}{1988}{`Computer Simulation
Using Particles'}{IOP Publishing}{ISBN 0-85274-392-0}
\paper{Jaroszy\'{n}ski M.}{1991}{MNRAS}{249}{430}
\paper{Jaroszy\'{n}ski M.}{1992}{MNRAS}{255}{655}
\paper{Jaroszy\'nski M., Park C., Paczynski B., \& Gott III J. R.}
{1990}{Ap. J.}{365}{22}
\paper{Keeton C. R. \& Kochanek C. S.}{1997}{Ap. J.}{487}{42}
\paper{Kovner I.}{1987}{Ap. J.}{316}{52}
\paper{Marri S. \& Ferrara A.}{1998}{astro-ph}{9806053}{Preprint}
\paper{Peacock J. A. \& Dodds S. J.}{1994}{MNRAS}{267}{1020}
\paper{Premadi P., Martel H. \& Matzner R.}{1998a}{Ap. J.}{493}{10}
\paper{Premadi P., Martel H. \& Matzner R.}{1998b}{astro-ph}
{9807127}{Preprint}
\paper{Premadi P., Martel H. \& Matzner R.}{1998c}{astro-ph}
{9807129}{Preprint}
\paper{Riess A. G., Filippenko A. V., Challis P., Clocchiatti A.,
Diercks A., Garnavich P. M., Gilliland R. L., Hogan C. J., Jha S.,
Kirshner R. P., Leibundgut B., Phillips M. M., Reiss D., Schmidt
B. P., Schommer R. A., Smith R. C., Spyromilio J., Stubbs C., Suntzeff
N. B. \& Tonry J.}{1998}{A. J.}{116}{1009}
\paper{Schneider P., Ehlers J., \& Falco E. E.}{1992} {`Gravitational
Lenses'}{Springer-Verlag}{ISBN 0-387-97070-3}
\paper{Schneider P. \& Weiss A.}{1988a}{Ap. J.}{327}{526}
\paper{Schneider P. \& Weiss A.}{1988b}{Ap. J.}{330}{1}
\paper{Thomas P. A., Webster R. L. \& Drinkwater
M. J.}{1995}{MNRAS}{273}{1069} 
\paper{Tomita K.}{1998}{astro-ph}{9806047}{Preprint}
\paper{Vianna P. T. P., \& Liddle A. R.}{1996}{MNRAS}{281}{323}
\paper{Wambsganss J., Cen R., \& Ostriker J.}{1998}{Ap. J.}{494}{29}
\paper{Wambsganss J., Cen R., Xu G. \& Ostriker
J.}{1997}{Ap. J.}{475}{L81}
\paper{Watanabe K.\& Tomita K.}{1990}{Ap. J.}{355}{1}

\end{document}